\documentclass[aps,twocolumn,prb,reprint,floatfix,showpacs,lengthcheck,citeautoscript,superscriptaddress,nofootinbib]{revtex4-2}
\usepackage{amsmath}
\usepackage{amssymb}
\usepackage{graphics}
\usepackage{graphicx}
\usepackage{bm}
\usepackage{float}
\usepackage{xcolor}
\usepackage{times}
\usepackage{notes2bib}
\usepackage{verbatim}
\usepackage[none]{hyphenat}
\usepackage{epstopdf} 
\usepackage{tikz}
\usepackage[colorlinks=true,
linkcolor=blue, %color for equations links
citecolor=blue, %color for cites links
urlcolor=blue,
bookmarks=true,
hypertexnames=true]{hyperref}

\newcommand{\ve}{\varepsilon}

\newcommand{\ci}{\mathrm{i}}

\newcommand{\ket}[1]{| #1 \rangle}
\newcommand{\bra}[1]{\langle #1 |}

\newcommand{\bigdot}{\boldsymbol{\cdot}}

\newcommand{\hw}{\hbar\omega}
\newcommand{\moi}{\leqslant}
\newcommand{\dk}{\frac{d\bm{k}}{(2\pi)^2}}

\newcommand{\e}{\epsilon}
\newcommand{\w}{\omega}
\newcommand{\z}{\zeta}
\newcommand{\p}{\partial}
\newcommand{\scp}{\scriptscriptstyle}
\newcommand{\chil}{\chi_{ij}^{\scriptscriptstyle (1)}}
\newcommand{\chiq}{\chi_{ijk}^{\scriptscriptstyle (2)}}
\begin{document}
	\title{Minimal $d$-Band Model for the Optical Susceptibility of Noncentrosymmetric Monolayer Transition Metal Dichalcogenides}
	\author{Angiolo Huamán}
	\affiliation{Independent researcher}
	\begin{abstract}
The optical response of two-dimensional materials has been customarily calculated {\it ab initio} using plane waves basis and the full Bloch wavefunction, without separating the most important orbital contributions. In the family of monolayer transition metal dichalcogenides  lacking inversion symmetry, we take advantage of the mostly $d-$orbital content of the Bloch bands around the semiconductor gap to reduce the calculation of their linear and quadratic optical susceptibilities to a very minimal model that includes only three energy bands. As the Bloch wavefunction determines the microscopic response to external fields, this simple approach reproduces well first principles calculations up to roughly 2 eV above the band gap. This could be the starting point for the inclusion of many-body effects with only a few energy bands in a numerically inexpensive way. 
	\end{abstract}
	\date{\today}
	\maketitle

\section{Introduction}\label{sec1}
The optical response of semiconductors; that is, the radiation fields emitted by them upon illumination, has been historically used to determine their structural~\cite{Pollak1990} and electronic properties, particularly their energy gaps~\cite{Macfarlane1955}. Even though nowadays we have more precise techniques, like ARPES~\cite{Damascelli2003,Zhang2022}, to characterize the electronic state of solids, the study of the optical response, both at the experimental and theoretical level, is still of a fundamental importance. 

Furthermore, monolayers of transition metal dichalcogenides (TMDCs)~\cite{Manzeli2017}, a family of two-dimensional (2D) semiconductors with band gap sizes in the optical region (around $1\,$eV), have attracted much attention because their intrinsic electronic properties invites both fundamental and applied investigations, with promising applications to optoelectronics (such as light-emitting devices~\cite{Pu2018}) and valleytronics~\cite{Xiao2012,Ominato2020}. Monolayer TMDCs also stand out as platforms for the study of many-body physics, as their low dimensionality enhances electron correlations, resulting in a wealth of excitonic states~\cite{Klein2021,You2015}. For many of these topics, an understanding of the response to an applied external driving field is essential.

Many different effects comprise the field of light-matter interaction, namely absorption by, transmission through, and reflection from a sample. In experiments dealing with the latter, the reflected patterns carry information about the structural properties of the material~\cite{Yao2021}. This is more evident when the {\it nonlinear} response is detected: it is nonzero only when the underlying crystal lacks an inversion center, while the linear response is present regardless of the crystal geometry and its symmetries.

The theoretical investigation of the emitted signal in reflection experiments is carried out at two levels: the determination of the reflected light from the oscillating polarization induced by the optical driving field in the sample, which amounts to solving Maxwell equations with the appropriate source and boundary conditions~\cite{Sipe1987GF}; and the actual calculation of the induced polarization, that is given by the {\it optical susceptibility}. The latter has been obtained---in noncentrosymmetric monolayer TMDCs---mostly by resorting to first principles methods, such as density functional theory (DFT) with a plane-wave basis~\cite{Pike2021,Wang2015}, supplemented by low-energy effective models, as those used to describe strain effects~\cite{Zollner2019}. Few-layer TMDCs have also been studied using DFT, and were shown to exhibit an important thickness dependence in their optical properties~\cite{Paul2022}.  However, DFT calculations reveal that the orbital content of the optical bands (those close to the energy gap) in {\it 2H} monolayer TMDCs mostly comes from the  {\it d}-orbitals of the transition metal, with negligible contributions from the {\it p}-orbitals of the chalcogen atoms. This fact poses the question of whether a calculation using a minimal orbital basis is feasible and how well it compares to more comprehensive approaches.

In this work, we build upon the three-band model with three transition metal {\it d}-orbitals by Liu {\it et al}~\cite{Liu2013} and calculate the linear and nonlinear (quadratic) optical susceptibility of monolayer TMDCs in the {\it 2H} stacking configuration~\cite{Sen2024} (i.e., lacking inversion symmetry), taking tungsten disulfide (WS$_2$) as an example, although the same calculation can be carried out for the entire family of TMDCs presented in~\cite{Liu2013}. This  is done in the single particle approximation, disregarding many-body effects, which can be important but difficult to include, particularly in the nonlinear response. As a consequence of an almost complete {\it d}-character of the energy bands around the semiconductor gap, momentum matrix elements are close to those obtained using the full DFT Kohn-Sham orbitals, and the optical susceptibility thus obtained  matches {\it ab initio} calculations for photon energies up to 1.7 eV above the band gap size. Effects due to spin-orbit coupling (SOC) are also included in an approximate way. Using this simplified model as a basis, many-body effects can be added by using techniques based on tight-binding (TB) Hamiltonians~\cite{Trolle2014}.

This paper is organized as follows. Section~\ref{sec2} gives a brief description of monolayer TMDCs in the {\it 2H} stacking. In Sec.~\ref{sec3} we present the formalism for the linear and second-order optical susceptibility in the single-particle approximation. The construction of the matrix elements of the momentum operator between Bloch eigenstates is described in Sec.~\ref{sec4}. We present the formalism for the calculation of two-center integrals of the momentum operator in Sec.~\ref{sec5}. Section~\ref{sec6} is an account of the symmetry considerations leading to the reduction of integrals over the Brillouin zone (BZ). Section~\ref{sec7} presents and discusses our results. The conclusions we arrived at are summarized in Sec.~\ref{sec8}.

\section{Monolayer transition metal dichalcogenides}\label{sec2}
Monolayers of TMDCs can have an allotropic form with a $D_{3h}$ point group, thus lacking an inversion center, as schematically shown in Figs.~\ref{crystal}(a) and (b). The lattice vectors are chosen as $\bm{a}_1=a\,\hat{\bm{x}}$ and $\bm{a}_2=a(\hat{\bm{x}}+\sqrt{3}\,\hat{\bm{y}})/2$, $a$ being the lattice parameter. The transition metal atoms occupy the Bravais lattice sites, while there are two chalcogens per primitive cell  at positions $(1/3)(\bm{a}_1+\bm{a}_2)\pm h\hat{\bm{z}}/2$, where $h$ is the chalcogen-chalcogen distance along the $z$ direction. The reciprocal lattice vectors are $\bm{b}_1=(4\pi/(a\sqrt{3}))(\sqrt{3}\,\hat{\bm{x}}-\hat{\bm{y}})/2$ and $\bm{b}_2=(4\pi/(a\sqrt{3}))\hat{\bm{y}}$, and the hexagonal Brillouin zone is depicted in Fig.~\ref{crystal}(c). The irreducible Brillouin zone (IBZ), defined as the part of the BZ which, upon application of all the symmetry operations in $D_{3h}$, generates the entire BZ, is also depicted as a gray region. The volume of the IBZ is the volume of the BZ divided by the order $g$ of $D_{3h}$ ($g=12$). We also show in Fig.~\ref{crystal}(d), as a hatched area, the so-called time-reversed irreducible Brillouin zone (tIBZ), which is just half the IBZ. It must be emphasized that the choices for both the IBZ and the tIBZ are not unique, although completely equivalent. These are all the properties needed for the formalism presented here, later on we will apply it in particular to tungsten disulfide WS$_2$.

The mirror plane at the $xy$ plane permits a decoupling between the two subsets of $d-$orbitals $\{d_{xy},d_{z^2},d_{x^2-y^2}\}$ and $\{d_{yz},d_{xz}\}$, with different parity with respect to the operation $\sigma_h$: $z\rightarrow -z$. Also,  {\it ab initio} calculations have  shown that the energy bands around the gap are mostly composed of orbitals of the former set, which allows the use of an effective three-band model as introduced by Liu {\it et al}~\cite{Liu2013}, whose details are given in Appendix~\ref{liuTB}.

Before going into the calculation of the optical susceptibility, in the next section we briefly review the main concepts and formulae.

\section{Optical susceptibility in the single particle approximation}\label{sec3}
We will work in the single-particle approximation and with an electromagnetic field with no spatial dependence (long wavelength approximation), described by the vector potential $\bm{A}(t)=\bm{A}(\w) e^{-\ci\omega t}+cc$. Although not explicitly written, $\w$ is to be understood as a complex frequency $\w+\ci\eta$, with $\eta$ a small positive quantity that guarantees that the vector potential $\bm{A}(t)$ vanishes at a remote past ($t\rightarrow-\infty$). The underlying monolayer TMDC is described by the static Hamiltonian $H_0=\bm{p}^2/2m+V(\bm{r})$ [$\bm{p}$ is the single particle momentum operator, $V(\bm{r})$ is the self-consistent ionic potential with the spatial symmetry of the crystal, and $m$ is the electron mass] with eigenvalues and Bloch eigenvectors satisfying $H_0\psi_{n\bm{k}}=\e_n(\bm{k})\psi_{n\bm{k}}$. In the dipole approximation, the time-dependent Hamiltonian $H(t)$ of the material coupled to the optical field is obtained using the Peierls substitution and keeping only  terms linear in the vector potential $\bm{A}(t)$:
\begin{figure}[t!]
	\centering
	\includegraphics[width=0.95\columnwidth]{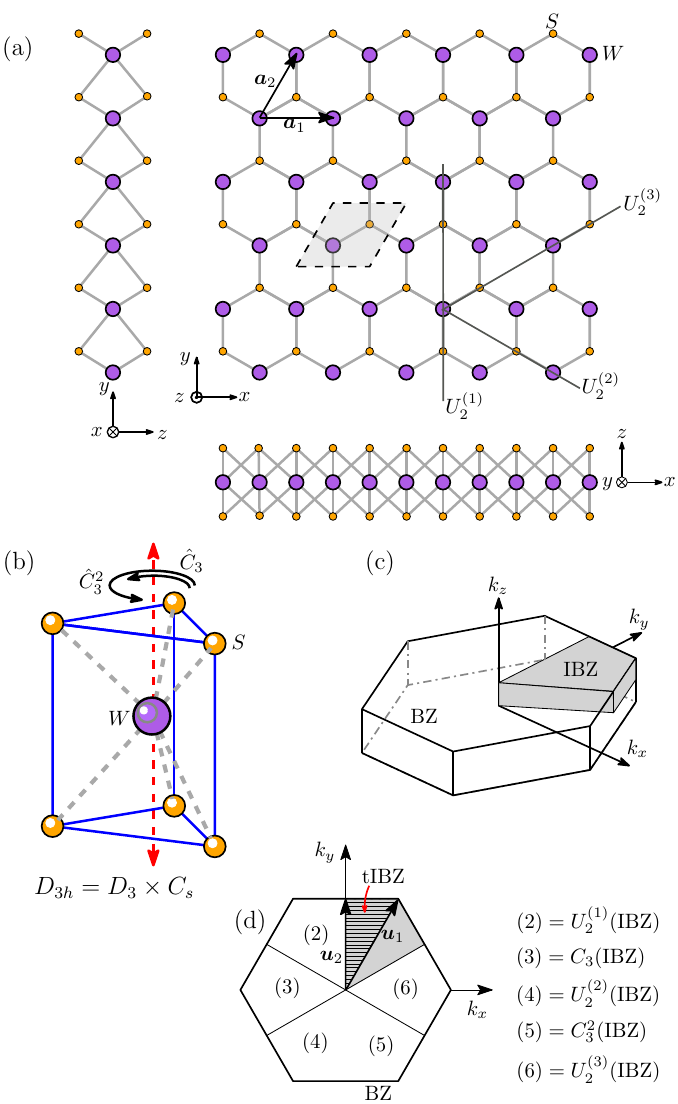}
	\caption{(a) Top view of monolayer WS$_2$ with trigonal prismatic configuration $2H$ and lattice vectors $\bm{a}_1$ and $\bm{a}_2$ (lattice parameter $a=|\bm{a}_1|=|\bm{a}_2|$). Sulfur layers lie one on top of the other. (b) Trigonal unit cell of WS$_2$ exhibiting its $D_{3h}$ symmetry. (c) Hexagonal first Brillouin zone (BZ), with the irreducible Brillouin zone (IBZ, gray region). (d) Two-dimensional time-reversed irreducible region (tIBZ, hatched area). Regions $(2)$ through $(6)$ are obtained from the IBZ by applying the symmetry operations indicated.\label{crystal}}
\end{figure}

\begin{align}\label{H0}
	H(t) =\big[\bm{p}+&e\bm{A}(t)\big]^2/2m+V(\bm{r})\notag\\
	     &\simeq H_0+H_1(t)+\mathcal{O}(A^2),
\end{align}
where $H_1(t)=(e/m)\bm{A}(t)\bigdot\bm{p}$ is the coupling with the optical field and  $-e$ is the electron charge.  The theory of the linear and nonlinear optical susceptibilities is abundant~\cite{Cabellos2009,Ghahramani1991,Sipe1993, Moss1986}, so in what follows we will summarize the main concepts and  equations only. 

\subsection{Linear optical susceptibility}
For monochromatic illumination, the single-body linear optical susceptibility tensor $\chi_{ij}^{(1)}(\w)$ in SI units is dimensionless and given in terms of the Fourier components of the induced polarization $\bm{P}(t)$ and the applied electric field $\bm{E}(t)$~\cite{Hemstreet1972,Cabellos2009}:
\begin{equation}
	P_i(\w)=\ve_0\chil(\w)E_j(\w),
\end{equation}
with $\bm{E}(\w)=\ci\w\bm{A}(\w)$ and $\ve_0$ is the vacuum permittivity. Throughout this paper, the Einstein sum convention over repeated Cartesian indices (denoted by $i$, $j$ and $k$, and their primed forms) will be used. The optical susceptibility is obtained by solving the density matrix in a perturbative way, from which the macroscopic induced current $\bm{J}(t)$ can be obtained. The macroscopic polarization is obtained through $\bm{P}(t)=\int_{-\infty}^t d\tau \bm{J}(\tau)$~\cite{JacksonCE}, up to the linear order in $\bm{A}(t)$, as it is shown elsewhere~\cite{Cabellos2009}. Considering only interband transitions we get the following expression:
%\begin{align}
%	{\chi}^{\scp (1)}_{ij}(\omega)=&\frac{e^2\hbar^2}{\ve_0 m^2} \frac{1}{V}\sum_{\bm{k}} \times \notag\\
%	&\sum_{v,c}\frac{p^i_{vc}p^j_{cv}}{\e^2_{cv}(\e_{cv}-\hw-\ci\eta)},\label{chi01}
%\end{align}
\begin{equation}\label{eq:chil1}
	{\chi}^{\scp (1)}_{ij}(\omega)=\frac{e^2\hbar^2}{\ve_0 m^2} \frac{1}{V}\sum_{\bm{k}}\sum_{v,c}\frac{p^i_{vc}p^j_{cv}}{\e^2_{cv}(\e_{cv}-\hw-\ci\eta)},
\end{equation}
with $\e_{cv}=\e_c-\e_v$, $p^i_{vc}=\bra{\psi_{v}}p^i\ket{\psi_{c}}$, $p^i$ is the $i-$th component of the  momentum operator, and $V$ is the volume of the sample. The imaginary part of $\w$ has been explicitly written, so that $\w$ is real in Eq.~\eqref{eq:chil1}. The $\bm{k}-$dependence on the energy bands and Bloch eigenstates is dropped not to load the notation. The sums run over valence ($v$) and conduction ($c$) Bloch eigenstates of $H_0$.  Following Ref.~\cite{Khan2022}, in the monolayer limit the Bloch wavefunctions are highly localized at $z=0$ (the plane of the monolayer), so the sum over $k_z$ can be replaced with $N_z$, where $N_z$ is the number of layers in the sample. If the interlayer distance is $a_3$, the volume $V$ can be written as $V=N_za_3 S$, where $S$ is the area of the layer. After passing to the continuous limit in the in-plane coordinates $k_x$ and $k_y$, we get:  
\begin{align}\label{eq:chil2}
	\chil(\omega)=\frac{1}{a_3}\frac{e^2\hbar^2}{\ve_0 m^2} &\int_\text{BZ}\dk \times \notag\\
	&\sum_{v,c}\frac{p^i_{vc}p^j_{cv}}{\e^2_{cv}(\e_{cv}-\hw-\ci\eta)},
\end{align}
where the integral is over the two-dimensional BZ only [see Fig.~\ref{crystal}(d)]. The expression in Eq.~\eqref{eq:chil2} can be rewritten by using time reversal symmetry, which implies that $(p^i_{vc}(\bm{k}))^*=-p^i_{vc}(-\bm{k})$, and allows to get its imaginary part by using the property $1/(x\pm\ci\eta)=\mathcal{P}(1/x)\mp\ci\pi\delta(x)$ ($\mathcal{P}$ is the Cauchy principal value~\cite{RileyMMPE}):
\begin{align}\label{chil11}
	\text{Im}[\chil(\omega)]=-&\frac{1}{a_3}\frac{\pi e^2\hbar^4}{\ve_0 m^2}\times\notag\\
	\sum_{v,c} &\int_\text{BZ}\dk\frac{\text{Re}[\z^i_{vc}\z^j_{cv}]}{\e^2_{cv}}\delta(\e_{cv}-\hbar\w).
\end{align}
In the expression above we have defined the matrix elements of the derivatives $\p/\p x_i$:
\begin{equation}\label{derivs}
\z^i_{cv}(\bm{k})=\int d\bm{r}\,\psi_{c\bm{k}}^*\frac{\partial}{\partial x_i}\psi_{v\bm{k}}.
\end{equation}
The Dirac delta in Eq.~\eqref{chil11} allows to pull the factor $1/\e_{cv}^2$ out of the integral and make it equal to $1/\w^2$. We have verified that the difference between both calculations lies within the numerical error. For this equation it might also be useful to keep in mind that $e^2\hbar^4/(\e_0m^2) \simeq 1.1\times10^4\,\mbox{eV}^3\mbox{\AA}^5$. The complex dielectric function is defined as $\ve_{ij}(\w)=\ve'_{ij}+\ci \ve{''}_{ij}=\delta_{ij}+\chil$, $\delta_{ij}$ being the Kronecker symbol. Once having $\chil$, related linear quantities such as the refractive index, reflectivity, and absorption spectrum can be calculated. The integration of Eq.~\eqref{chil11} has been accomplished by a {\it broadening} method consisting in replacing $\delta(\e_{cv}-\hbar\w)$ with a weighted  Gaussian function of a small width~\cite{Guo2005}. This approach results in  a smoother dispersion of the susceptibility at the expense of a denser integration mesh, compared to other approaches, such as the linear tetrahedral method~\cite{Lehmann1972}. In Sec.~\ref{sec7} we give more details on our integration procedure.

\subsection{Second order susceptibility}

The second order susceptibility tensor $\chiq$ is defined in S.I. units as follows~\cite{Boyd,Zu2022}:
\begin{equation}\label{eq:P2w}
	P_i(\w)=\e_0 \chiq(\w) E_j(\w) E_k(\w).
\end{equation}
From this definition it follows that $\chiq$ is symmetric in its last two indices, $\chi_{ijk}^{\scriptscriptstyle (2)}=\chi_{ikj}^{\scriptscriptstyle (2)}$, and that has dimensions of length over voltage. The expression for $\chiq$ in terms of eigenstates and energy bands of $H_0$ is obtained in a similar manner as for $\chil$, this time by expanding the density matrix up to second order in the vector potential $\bm{A}(t)$. This procedure is rather lengthy but has been explored abundantly in the literature~\cite{Cabellos2009,Ghahramani1991,Sipe1993, Moss1986}, so here we will present the final formulae only.

It is convenient to write $\chiq$ as a sum of two terms, $\chiq=A_{ijk}+B_{ijk}$, each one symmetric in their last two indices and with imaginary parts given by~\cite{Cabellos2009}:
\begin{widetext}
	\begin{align}
	\mbox{Im}[A_{ijk}] &=\frac{1}{a_3}\frac{e^3\hbar^6}{2\ve_0m^3}\int \dk \sum_{v,c} \frac{16\pi}{\e_{cv}^3}\delta(\e_{cv}-2\hbar\w) \Bigg( 
	\sum_{v'}\frac{\mbox{Im}[\ci\z^i_{vc}\{\z^j_{cv'},\z^k_{v'v}\}]}{2\e_{cv'}-\e_{cv}} -
	\sum_{c'}\frac{\mbox{Im}[\ci\z^i_{vc}\{\z^j_{cc'},\z^k_{c'v}\}]}{2\e_{c'v}-\e_{cv}} \Bigg), \label{shg11}\\
	\mbox{Im}[B_{ijk}] &=\frac{1}{a_3}\frac{e^3\hbar^6}{2\ve_0m^3}\int \dk \sum_{v,c} \frac{\pi}{\e_{cv}^3}\delta(\e_{cv}-\hbar\w) \Bigg( 
	\sum_{n\neq c}\frac{\mbox{Im}[\ci\z^i_{nc}\{\z^j_{cv},\z^k_{vn}\}]}{\e_{cn}-2\e_{cv}} -
	\sum_{n\neq v}\frac{\mbox{Im}[\ci\z^i_{vn}\{\z^j_{nc},\z^k_{cv}\}]}{\e_{nv}-2\e_{cv}} \Bigg). \label{shg22}
	\end{align}
\end{widetext}
In these expressions we have introduced the {\it symmetrizing} operator $\{R^j,L^k\}=(R^jL^k+R^kL^j)/2$, that enforces the symmetry $\chi^{\scp (2)}_{ijk}=\chi^{\scp (2)}_{ikj}$. It might be useful to remember that $e^3\hbar^6/(\e_0m^3) \simeq 8.1\times10^4\,\text{\AA}^7\text{eV}^5/\mbox{V}$. As with the linear susceptibility, $\chiq$ has been scaled with the interlayer distance by the factor $1/a_3$, while the integration covers the 2D BZ. Also, the separation into Eqs.~\eqref{shg11} and~\eqref{shg22} is more than just formal: the different Dirac deltas in Eq.~\eqref{shg11} and~\eqref{shg22} are treated differently when doing the integration over the Brillouin zone. Similarly to the linear susceptibility, the denominators $\e_{cv}^3$ are constant over the surface of integration determined by the Dirac deltas and can be pulled out of the integral.

Both Eq.~\eqref{chil11} and Eqs.~\eqref{shg11} and~\eqref{shg22} depend on momentum matrix elements and an integration over the whole BZ in their definitions. In the next two sections we first construct these matrix elements and then present the symmetry arguments leading to the integration over a reduced part of the BZ: the tIBZ.

    As a side note, it must be emphasized that the equations above are valid in the single-body approximation, where electron-electron Coulomb interactions are {\it averaged out} and included in the ionic background potential $V(\bm{r})$ in a self-consistent manner (as it is done in most DFT implementations). On the other hand, the inclusion of many-body correlations entails the formation of excitons with discrete energies within the single-particle semiconductor gap, and which appear as sharp features in photoluminescence experiments~\cite{Ugeda2014}. The energy of the lowest excitonic state with respect to the single-body conduction band minimum in WS$_2$ varies depending on the substrate and experimental technique, ranging from 2.31 eV~\cite{Stier2016} to 3.01 eV~\cite{Hanbicki2015}. These many-body effects are enhanced by the low dimensionality of monolayer TMDCs~\cite{Cudazzo2016}. However, the starting point for the calculation of many-body corrections is most commonly the single-body wavefunctions, chiefly using DFT. From them, the exciton wavefunction $\Psi_x$ can be expanded as a tensor product of valence ($v$) and conduction ($c$) one-particle eigenstates~\cite{Glazov2015}:
    \begin{equation}
        \Psi_x=\sum_{v,c}\sum_{\bm{k}_v,\bm{k}_c} C_{vc}(\bm{k}_v,\bm{k}_c)\psi_{v\bm{k}_v}\otimes\psi_{c\bm{k}_c},
    \end{equation}
    where the coefficients $C_{vc}$ are obtained by solving the Bethe-Salpeter equation, for instance. Thus, apart from a renormalization of the energy gap size and an enhancement because of the electron-electron correlations, the experimental optical response is the single-body one plus sharp peaks coming from the discrete excitonic energy levels.  Having  said this, in what follows we will focus entirely on the single-body optical response.

\section{Symmetry of the Bloch eigenstates as combinations of atomic pseudo orbitals}\label{sec4}
Equations~\eqref{chil11}, \eqref{shg11} and~\eqref{shg22} contain matrix elements of the momentum operator between Bloch eigenstates, which have not been constructed yet. In this section we address that point. These Bloch eigenstates for the non illuminated monolayer TMDC can be expanded, as an approximation, in a basis of pseudo atomic orbitals (PAOs) obtained from DFT. Liu {\it et al}~\cite{Liu2003} have shown that the broad family of  monolayer TMDCs with formula MX$_2$  (M=W, Mo and X=S, Se, Te ) in the {\it 2H} configuration [i.e., without an inversion center, see Fig.~\ref{crystal}(b)] can be described by using only some {\it d}-orbitals of the transition metal if only the bands around the semiconductor gap (which are the most relevant for optical excitations) need to be correctly described. Moreover, because of the $\sigma_h$ symmetry present in their $D_{3h}$ point group, only {\it d}-orbitals even in {\it z}, that is, $d_{xy}$, $d_{x^2-y^2}$ and $d_{z^2}$, have to be included. Density functional theory calculations with the VASP implementation using the projector-augmented wave method show that the orbital content of the three bands closest to the energy gap have precisely a {\it d}-character~\cite{Liu2003}. For each PAO $(d_1,d_2,d_3)=(d_{z^2},d_{xy}$, $d_{x^2-y^2})$ of the transition metal in the primitive cell, the corresponding Bloch sum can be written as follows:
\begin{equation}\label{sums}
	\varphi_{n\bm{k}}(\bm{r})=\frac{1}{\sqrt{N}}\sum_{\bm{R}} e^{\ci \bm{k}\bigdot \bm{R}} d_n(\bm{r}{-}\bm{R}),\;n=1,2,3,
\end{equation}
where the sum is over the $N$ lattice sites $\bm{R}=\sum_jn_j\bm{a}_j$ ($n_j$ integers) making up the sample. As customary, the matrix elements of the TB Hamiltonian $H_0$ matrix are taken between these Bloch sums $\varphi_{n\bm{k}}(\bm{r})$:
\begin{equation}\label{HTB}
	H_{0,ab}(\bm{k})=\int d\bm{r}\,\varphi^*_{a\bm{k}}(\bm{r})H_0(\bm{r})\varphi_{b\bm{k}}(\bm{r}).
\end{equation}
If we restrict $\bm{k}$ in Eq.~\eqref{HTB} to be in the IBZ, $H_{0,ab}(\bm{k})$ for other $\bm{k}$ points can be directly obtained. To this, let $\hat{G}$ be a symmetry operation belonging to the group $D_{3h}$, and let us call the matrix performing the corresponding coordinate transformation $\tilde{G}$:
	\begin{equation}\label{eq:Gtilde}
		x'_i=\tilde{G}_{ij}x_j,
	\end{equation} 
where $x_i'$ are the coordinates after the transformation. Performing the substitution in Eq.~\eqref{eq:Gtilde} in the integral in Eq.~\eqref{HTB} it can be shown that:
\begin{equation}\label{HG}
	H_{0,ab}(\hat{G}\bm{k})=\sum_{\mu,\nu}G_{\mu a}H_{0,\mu\nu}(\bm{k})G_{\nu b}, 
\end{equation} 
where $G_{\mu\nu}$ are the entries of the orthogonal representations matrix of $\hat{G}$ in the orbital basis $\{d_1,d_2,d_3\}=\{d_{z^2},d_{xy},d_{x^2-y^2}\}$ introduced above, that is, $d_\alpha(\hat{G}\bm{r})=\sum_\beta G_{\beta\alpha} d_\beta(\bm{r})$. This representation is reducible and can be written as $A_1^\prime+E^\prime$~\cite{TinkhamGTaQM}. The form of the $G$ matrices can be worked out by noticing that $D_{3h}=D_{3}\times C_s$~\cite{LandauQMNRT}, where $C_s$ is the point group containing only the identity and the reflection $\sigma_h$ in the $xy$ plane (the plane of the monolayer). This operation has the only effect of changing $z$ with $-z$, and becomes redundant in our two-dimensional representation of $\bm{k}-$space. Since the full 2D BZ can be covered using the $D_{3}$ point group, only the representations of the latter will be necessary. It is clear that all of the operations in this group satisfy $d_{z^2}(\hat{G}\bm{r})=d_{z^2}(\bm{r})$. Since the symmetry operations of the point group $D_{3d}$ can only keep the coordinate $z$ or reverse it, their corresponding matrices for coordinate transformations [Eq.~\eqref{eq:Gtilde}] are:
\begin{equation}\label{Gmatrix}
	\tilde{G}=\left(\begin{array}{ccc}
		g_{11} & g_{12} & 0 \\
		g_{21} & g_{22} & 0 \\
		0      & 0      & \pm1
	\end{array}\right),
\end{equation}
and the representations matrices in the basis $\{d_1,d_2,d_3\}=\{d_{z^2},d_{xy},d_{x^2-y^2}\}$ are:
\begin{equation}
	G=\left(\begin{array}{ccc}
		1 & 0                         & 0                \\
		0 & g_{11}g_{22}+g_{12}g_{21} & 2g_{11}g_{12}    \\
		0 & 2g_{11}g_{21}             & g_{11}^2-g_{21}^2 
	\end{array}\right).
\end{equation}
With the expression above, Eq.~\eqref{HG} can be written in matrix form as $H_0(\hat{G}\bm{k})=G^T H_0(\bm{k})G$. Since every point in the BZ can be obtained from another one in the IBZ upon application of a suitably chosen $\hat{G}$ operation, the TB Hamiltonian matrix given in Eq.~\eqref{HTB}  have to be numerically diagonalized only for points in the IBZ. Time reversal symmetry allows for a further reduction to the tIBZ. The specific form of $H_{0,ab}(\bm{k})$ depends on the details of the TB model used. The approximation we will use includes third-nearest neighbors, as explained later on and in Appendix~\ref{liuTB}.
 
Let us pass to the calculation of the matrix elements of the momentum operator. Customarily, these have been based on the projector-augmented wave (PAW) method~\cite{Adolph2001}. In our case, the Bloch eigenstates of $H_0$ (see Sec.~\ref{sec3}) in the $\bm{r}-$representation will be expanded as a combination of the Bloch orbitals $\varphi_{s\bm{k}}(\bm{r})$ already defined:
\begin{equation}\label{blochpsi}
	\psi_{n\bm{k}}(\bm{r})=\sum_s C_{sn}(\bm{k}) \varphi_{s\bm{k}}(\bm{r}),
\end{equation}
the matrix elements $C_{sn}(\bm{k})$ are obtained by diagonalizing the  TB Hamiltonian Eq.~\eqref{HTB} with DFT parameters, resulting in $H_0(\bm{k})C(\bm{k})=C(\bm{k})\e(\bm{k})$, where $\e(\bm{k})$ is a diagonal matrix with $\e_{ab}(\bm{k})=\e_a(\bm{k})\delta_{ab}$, containing the energy bands $\e_a(\bm{k})$. This relation, together with Eq.~\eqref{HTB}, allow us to choose the $gauge$ $C(\hat{G}\bm{k})=G^T C(\bm{k})$, for $\bm{k}$ in the IBZ. Moreover, with this choice it is guaranteed the well known property in Bloch theory~\cite{KittelQTS}:
\begin{equation}
	\psi_{n\bm{k}}(\hat{G}\bm{r})=\psi_{n\hat{G}^{-1}\bm{k}}(\bm{r}),
\end{equation}
for non-degenerate energy bands. Using the definition in Eq.~\eqref{derivs}, the matrix elements $p^i_{ab}(\bm{k})$ of the momentum operator are more conveniently written as $p^i_{ab}(\bm{k})=(\hbar/\ci)\z_{ab}^i(\bm{k})$, where $\z_{ab}^i(\bm{k})$ are matrix elements of the derivative operator $\partial/\partial x^i$ [Eq.~\eqref{derivs}] and can be written using Eqs.~\eqref{sums} and~\eqref{blochpsi} as follows: 
\begin{equation}\label{matrix_elements}
	\z^{(j)}_{ab}(\bm{k})=\sum_{s,s'}C^*_{sa}(\bm{k})C_{s'b}(\bm{k})\sum_{\bm{R}}e^{-\ci\bm{k}\bigdot\bm{R}}D^{(j)}_{ss'}(\bm{R}).
\end{equation}
The quantity $D^{(j)}_{ss'}(\bm{R})$ is a two-center integral of the derivative operator between localized PAOs separated by $\bm{R}$:
\begin{equation}\label{d_integrals}
  D^{(j)}_{ss'}(\bm{R})=\int d\bm{r}\,d_s(\bm{r}{-}\bm{R})\frac{\p}{\p x^j} d_{s'}(\bm{r}).
\end{equation}
Integrating by parts the equation above and keeping in mind the parity of the {\it d}-orbitals, it is easy to verify its symmetry, i.e., 
\begin{equation}\label{Dsymms}
	D^{(j)}_{ss'}(\bm{R})=D^{(j)}_{s's}(\bm{R}).
\end{equation}
It might be useful to notice that, defining matrices $\z^{(j)}(\bm{k})$ and $D^{(j)}(\bm{R})$ with entries $\z^{(j)}_{ab}(\bm{k})$ and $D^{(j)}(\bm{R})_{ss'}$ respectively, Eq.~\eqref{matrix_elements} can be written as $\z^{(j)}(\bm{k})=\sum_{\bm{R}} e^{-\ci \bm{k}\bigdot\bm{R}}C^\dagger(\bm{k})D^{(j)}(\bm{R})C(\bm{k})$. 

Before moving forward, we must state the transformation rules of $\z^{(j)}_{ab}(\bm{k})$ under $D_{3h}$ symmetries, as these will be needed for the reduction of integrals in the BZ. Replacing $\bm{k}$ with $\hat{G}\bm{k}$ ($\hat{G}$ belonging to $D_{3h}$) in Eq.~\eqref{matrix_elements}, and using the facts that $C(\hat{G}\bm{k})=G^TC(\bm{k})$ and $d_\alpha(\hat{G}\bm{r})=G_{\beta\alpha} d_\beta(\bm{r})$, it can be shown that: 
\begin{equation}\label{ztransf}
\z^{(j)}_{\mu\nu}(\hat{G}\bm{k})=\tilde{G}_{ji}\z^{(i)}_{\mu\nu}(\bm{k})
\end{equation}
with $\tilde{G}_{ji}$  defined in Eq.~\eqref{eq:Gtilde}. To completely determine the matrix elements $\zeta^{(j)}_{ab}(\bm{k})$ we need to calculate the two-center integrals $D^{(j)}_{ss'}(\bm{R})$, as we do in Sec.~\ref{sec5}. Before proceeding to that, we will show how SOC is introduced in the previous formalism.

\subsection{Spin-orbit effects}
It is well known that, contrary to graphene and because of the high atomic number in the transition metal, SOC is important in monolayer TMDCs~\cite{Zollner2025}. In particular, the tungsten-based family can have valence band splittings of almost $0.4\,$eV at the $K$ points, with a significantly smaller splitting at the bottom of the conduction band~\cite{ChiSinTang}. In this section we address how SOC can be included in our approach. Following Liu {\it et al}, the three-band {\it d}-orbital approximation reduces the $\bm{S}\bigdot\bm{L}$ SOC term to the $z$-direction only, as explained in the Appendix~\ref{liuTB}. Using the spin-orbital basis $\{\bar{d}_1,\cdots,\bar{d}_6\}=\{d_{z^2\uparrow},d_{xy\uparrow}$, $d_{x^2-y^2\uparrow}, d_{z^2\downarrow},d_{xy\downarrow}$, $d_{x^2-y^2\downarrow} \}$, the corresponding Bloch sums follow from Eq.~\eqref{sums}:
\begin{equation}\label{eq:spin_sums}
	\bar{\varphi}_{s\bm{k}}(\bm{r})=\frac{1}{\sqrt{N}}\sum_{\bm{R}} e^{\ci \bm{k}\bigdot \bm{R}} \, \bar{d}_s(\bm{r}{-}\bm{R}),\;s=1,\cdots,6.
\end{equation}
 In the Bloch basis above the Hamiltonian matrix is block diagonal and given by Eq.~\eqref{eq:HSOC} in Appendix~\ref{liuTB}. Spin-orbit coupling affects the optical response chiefly via two mechanisms. For one thing, the velocity operator~\cite{misc1} must now be obtained as the commutator of the position operator and the full Hamiltonian $H_\text{SOC}$ (including SOC)~\cite{Kim2018}:
\begin{eqnarray}\label{eq:gen_v}
    \bm{v}=\frac{1}{\ci\hbar}[\bm{r},H_\text{SOC}]=\frac{\bm{p}}{m}+\bm{v}_a,
\end{eqnarray}
where $\bm{p}$ is the momentum operator and $\bm{v}_a$ is an anomalous velocity coming from terms in $H_\text{SOC}$ that depend on the orbital angular momentum.
 Thus, changes in the optical susceptibility, as expressed by Eqs.~\eqref{chil11}, \eqref{shg11} and \eqref{shg22}, come from two main sources: (i) the modified momentum matrix elements, which are no longer simply proportional to the partial derivatives because of Eq.~\eqref{eq:gen_v}, and (ii)  the modification of the energy bands and Bloch eigenstates because of SOC.  For the family of monolayer TMDCs, first principle calculations using fully relativistic pseudopotentials have shown that the first effect is the order of less than 1\% of the zero-SOC response~\cite{Kim2018}, so will disregard it in this work. 

By including the spin degree of freedom the Bloch eigenfunctions are now written as:
\begin{equation}\label{blochpsiSOC}
	\psi_{s\bm{k}}(\bm{r})=\sum_{s'} C_{ss'}(\bm{k}) \bar{\varphi}_{s'\sigma\bm{k}}(\bm{r}).
\end{equation}
The coefficients $C_{ss'}(\bm{k})$ are obtained by numerical diagonalization of the Hamiltonian $H_\text{SOC}(\bm{k})$ in Eq.~\eqref{eq:HSOC}, Appendix~\ref{liuTB}. Clearly, the two-center integrals in Eq.~\eqref{d_integrals} are unaffected by the inclusion of the spin. Also, the orthogonality between $\bar{d}_s$ orbitals  corresponding to different spins is automatically accounted for by the block form of $H_\text{SOC}(\bm{k})$.

\section{Two-center integrals of the momentum operator}\label{sec5}
The numerical evaluation of two-center integrals of the like of Eq.~\eqref{d_integrals} is a recurrent problem in DFT and related areas. Similar to the Slater-Koster approach~\cite{Slater1954}, we combine the intrinsic symmetries of  the {\it d}-orbitals upon orthogonal transformations with an axes rotation to bring Eq.~\eqref{d_integrals} into a combination of just a few numerically evaluated integrals.

We perform an axes rotation that brings $\bm{R}$ into the $\hat{\bm{x}}$ direction. In this configuration, the integrals $D^{(j)}_{ss'}(a_p\hat{\bm{x}})$ ($a_p$ is the magnitude of $\bm{R}$, that will depend on whether we are dealing with nearest, second nearest or third nearest atomic neighbors) are not all independent, but only a few of them have to be numerically evaluated. The final values of Eq.~\eqref{d_integrals} are obtained by bringing back the $x$ axis to $\bm{R}$ and using the fact that PAOs corresponding to the same angular momentum $l$ transform into each other (they make a representation of the continuous group $C_\infty$ of arbitrary rotations around the $z$ axis). This approach reduces the computational effort by minimizing the number of numerical integrations. 
 
\begin{figure}[t!]
	\centering
	\includegraphics[width=0.9\columnwidth]{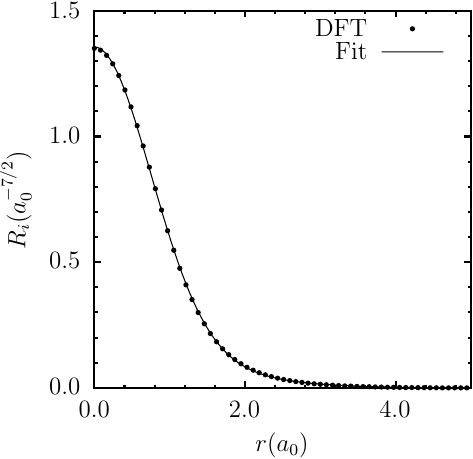}
	\caption{Radial part of the {\it d} PAO for tungsten. The dots are the DFT data while the solid line is the fitting in Eq.~\eqref{fit} ($a_0$ is the Bohr radius).\label{fig:PAO}}
\end{figure}

In the monolayer limit, these rotations are only in-plane, so the general vector $\bm{R}$ can be written as $\bm{R}=a_p(\cos\phi\,\hat{\bm{x}}+\sin\phi\,\hat{\bm{y}})$, the matrix performing this rotation is:
\begin{equation}\label{rot}
	P(\phi)=\left(\begin{array}{ccc}
		\cos\phi & -\sin\phi & 0\\
		\sin\phi &  \cos\phi & 0\\
		0 & 0 & 1 
	\end{array} \right).
\end{equation}
%Upon this transformation the partial derivatives go into:
%\begin{equation}
%	\frac{\partial}{\partial x'_j} \rightarrow M_{jk}(\bm{R})\frac{\partial}{\partial x_k},
%\end{equation}
For a given value of $a_p$, only the matrix elements $D^{(1)}_{11}(a_p\hat{\bm{x}})$, $D^{(1)}_{22}(a_p\hat{\bm{x}})$, $D^{(1)}_{33}(a_p\hat{\bm{x}})$, $D^{(1)}_{13}(a_p\hat{\bm{x}})$, $D^{(2)}_{12}(a_p\hat{\bm{x}})$ and $D^{(2)}_{23}(a_p\hat{\bm{x}})$ are needed, supplemented with the identity Eq.~\eqref{Dsymms}. Because of the $\sigma_h$ symmetry, all $D^{(3)}_{ss'}(a_p\hat{\bm{x}})$ are zero, which in turn will result in $\chil=0$ and $\chiq=0$ if at least one of their indices is $z$. The general integral in Eq.~\eqref{d_integrals} can thus be written as (sums run over all repeated indices):
\begin{align}\label{eq:Df}
	D^{(j)}_{ss'}(\bm{R})=P_{d,as} &P_{ji}P_{d,bs'}\times\notag\\
	&\int d\bm{r}\,d_a(\bm{r}{-}a_p\hat{\bm{x}})\frac{\p}{\p x^i}d_{b}(\bm{r}),
\end{align}
where $P_d$ is the representation matrix of the the {\it d}-orbitals $\{d_1,d_2,d_3\}$ under the rotation in Eq.~\eqref{rot}:
\begin{equation}\label{rotd}
	P_d(\phi)=\left(\begin{array}{ccc}
		1 &   0        &   0        \\
		0 &  \cos2\phi &  \sin2\phi \\
		0 & -\sin2\phi &  \cos2\phi 
	\end{array} \right),
\end{equation}
that is $d_\alpha(P\bm{r})=\sum_\beta P_{d,\beta\alpha}d_\beta(\bm{r})$. The $\phi$ dependence of $P$ and $P_d$ in Eq.~\eqref{eq:Df} has been omitted for clarity. The radial part of the {\it d}-orbitals is obtained from DFT calculations using the SIESTA package~\cite{Soler2002}, and it is written as $R(r)=r^2R_i(r)$, where $R_i$ is given by the SIESTA output as a table, which we verified that can be fitted into combinations of Gaussian functions. For WS$_2$, there is only one {\it d}-orbital channel and thus only one radial part. We verified that two Gaussians were sufficient:
\begin{equation}\label{fit}
	R_i(r)=a_1 e^{-b_1r^2}+a_2 e^{-b_2r^2},
\end{equation}
with parameters $a_1=0.135064\,a_0^{-7/2}$, $b_1=0.244444\,a_0^{-2}$, $a_2=1.22115\,a_0^{-7/2}$ and $b_2=0.904556\,a_0^{-2}$ , fitted from DFT PAOs ($a_0$ is the Bohr radius). The fitted equation is normalized to 0.001\%, so we renormalize it again. The form of the Eq.~\eqref{fit} has the advantage that $(1/r)dR_i/dr$ is well behaved at the origin, which simplifies the numerical implementation of the derivatives in Eq.~\eqref{d_integrals}. Figure~\ref{fig:PAO} shows the DFT data (dots) and the fitting in Eq.~\eqref{fit} (solid line). 

For the few integrals that have to be numerically calculated, we used the Gauss-Legendre quadrature over a box centered at $(a_p/2,0,0)$ ($a_p=|\bm{R}|$) and with dimensions $(a_p+2R_c)\times 2R_c\times 2R_c$, where $R_c$ is slightly larger than the maximum radial extension of the PAOs (in our case, roughly 4~\AA, corresponding to the {\it d}-orbitals of tungsten; see Fig.~\ref{fig:PAO}).

\section{Nonzero components of $\chil$ and $\chiq$}\label{sec6}

As it is well known, the point group of a crystal (its crystal class), greatly constrains the nonzero elements of its optical susceptibility at any order~\cite{Boyd,Shen}. For the linear susceptibility, this is most easily seen by means of the substitution $\bm{k}\rightarrow\hat{G}\bm{k}$ in Eq.~\eqref{chil11}, which does not affect either the size or the shape of the BZ, since the latter has all the symmetries of the $D_{3h}$ point group. Using the transformation in Eq.~\eqref{ztransf} we get:
\begin{equation}
	\chil=\tilde{G}_{ii'}\tilde{G}_{jj'}\chi_{i'j'}^{\scp (1)}.
\end{equation}
This  constitutes, for each $\hat{G}$, a homogeneous system of six linear algebraic equations (because of the symmetry $\chil=\chi^{\scp (1)}_{ji}$), and we have a similar system for each  symmetry operation in $D_{3h}$. Upon consecutive application of all the symmetry operations in $D_{3h}$ it can be verified the following [$x$, $y$ and $z$ subscripts indicate Cartesian coordinates according to Fig.~\ref{crystal}(a)]:
\begin{equation}\label{chixx}
	\chi_{xx}^{\scp (1)}=\chi_{yy}^{\scp (1)},
\end{equation}
all other entries being zero. The same symmetries, along with time reversal, help us reduce the integral over the BZ to only the tIBZ [see Fig.~\ref{crystal}(d)]. To this end, we follow a symmetrizing strategy~\cite{Chadi1973}. For the linear susceptibility, the integral in Eq.~\eqref{chil11} can be written in the following fashion:
\begin{align}\label{chi1symm}
	\text{Im}[\chil(\omega)]&=-\frac{1}{a_3}\frac{\pi e^2\hbar^4}{\ve_0 m^2}\sum_{v,c} \int_{\text{BZ}}\dk\times\notag\\
	&\sum_{\hat{G}}\text{Re}[\z^i_{vc}(\hat{G}\bm{k})\z^j_{cv}(\hat{G}\bm{k})]/g]\frac{\delta(\e_{cv}-\hbar\w)}{\e^2_{cv}},
\end{align}
where the sum over $\hat{G}$ extends to all the symmetry operations in the point group $D_3$ (the $C_s$ part in the decomposition $D_{3h}=D_3\times C_s$ becomes redundant in our 2D approximation) and $g=6$ is the order of $D_3$. The substitution $\bm{k}\rightarrow\hat{G}\bm{k}$ is  indicated only in the momentum matrix elements as the energy bands already have the symmetry $\e_{cv}(\bm{k})=\e_{cv}(\hat{G}\bm{k})$. In this way, the integrand in Eq.~\eqref{chi1symm} has the full $D_3$ symmetry, and thus can be reduced to an integral over the IBZ times $g$. Time reversal symmetry allows for a further reduction to the tIBZ times two. Thus, Eq.~\eqref{chi1symm} goes over into:
\begin{align}\label{chi1symm2}
	\text{Im}[\chil &(\omega)]=-\frac{2}{a_3}\frac{\pi e^2\hbar^4}{\ve_0 m^2}  \sum_{v,c}\int_{\text{tIBZ}}\dk \times\notag\\
	&\frac{1}{\e^2_{cv}}\sum_{\hat{G}}\text{Re}[\z^i_{vc}(\hat{G}\bm{k})\z^j_{cv}(\hat{G}\bm{k})] \, \delta(\e_{cv}-\hbar\w),
\end{align}
By using Eq.~\eqref{ztransf}, the integral in the equation above can be written as combinations of integrals of the kind in Eq.~\eqref{chil11} but over the tIBZ. More precisely, this can be done by using the transformation law:
\begin{equation}
	\z^i_{vc}(\hat{G}\bm{k})\z^j_{cv}(\hat{G}\bm{k})=\tilde{G}_{ii'}\tilde{G}_{jj'}\z^{i'}_{vc}(\bm{k})\z^{j'}_{cv}(\bm{k})
\end{equation} 
By means of this, $\chi^{\scp (1)}_{xx}$ can be written as a combination of integrals over the tIBZ only :
\begin{align}\label{chilibz}
	\text{Im}[\chi^{\scp (1)}_{xx}(\w)]&=-\frac{6}{a_3}\frac{\pi e^2\hbar^4}{\ve_0 m^2}\Bigg[ \notag\\
	&\int_\text{tIBZ}\dk\sum_{v,c}\frac{\z^x_{vc}\z^x_{cv}}{\e^2_{cv}}\delta(\e_{cv}-\hw)+ \notag\\
	&\int_\text{tIBZ}\dk\sum_{v,c}\frac{\z^y_{vc}\z^y_{cv}}{\e^2_{cv}}\delta(\e_{cv}-\hw)\Bigg],
\end{align}
Exactly the same equation holds true for $\chi^{\scp (1)}_{yy}$, as it should be. All other components are zero and $\chi^{\scp (1)}_{zz}$ is also so because of the mirror plane  passing through the transition metal atoms. 

While multilayer TMDCs (with an even number of layers)  do not have a second-order nonlinear susceptibility because of its $D_{6h}$ point group, which contains inversion as one of its symmetries, monolayer TMDCs  (with $2H$ configuration) have a reduced $D_{3h}$ point group with no inversion operation. For the second order susceptibility a similar result holds, namely:
\begin{equation}
	\chiq = \tilde{G}_{ii'} \tilde{G}_{jj'} \tilde{G}_{kk'}\chi_{i'j'k'}^{\scp (2)}.
\end{equation}
Because of the $\sigma_h$ mirror plane, all the components with at least a $z$ index are zero. Applying all the other symmetries we get:
\begin{equation}\label{eq:nzchiq}
	\chi^{\scp (2)}_{xxy}=\chi^{\scp (2)}_{yxx}=-\chi^{\scp (2)}_{yyy}.
\end{equation}
Similarly to the linear case, the integrand in Eqs.~\eqref{shg11} and~\eqref{shg22} can be symmetrized, thus reducing the integral over the whole BZ to the tIBZ. Without going into details, the final result is:
\begin{equation}\label{chiqibz}
	\chi^{\scp (2)}_{xxy}=2\big[3\bar{\chi}^{\scp (2)}_{xxy}+\frac{3}{2}(\bar{\chi}^{\scp (2)}_{yxx}-\bar{\chi}^{\scp (2)}_{yyy})\big],
\end{equation}
where the bars over the expressions on the right-hand of the equation above mean that they are to be calculated using Eqs.~\eqref{shg11} and~\eqref{shg22} {\it but with an integration over the tIBZ only}. It must be emphasized that Eqs.~\eqref{chilibz} and~\eqref{chiqibz} are in no way particular to monolayer TMDCs, but follow directly from the $D_{3h}$ symmetry and thus apply equally to any 2D material belonging to this crystal class.

    Before finishing this section, it might be useful to mention the way in which the symmetry of the optical susceptibility can be revealed in experimental setups~\cite{Lupke1999}. We will focus on SHG, which is the most important for crystal characterization.  The high-harmonic (frequency $2\w$) reflected field is expressed in terms of the incident field (frequency $\w$), and this relation is fixed by the nonzero and independent entries of the $\chiq$ tensor~\cite{Sipe1987}, which are set purely from symmetry grounds, and are independent of the microscopic model used. The induced electrical polarization $\bm{P}$ in the sample is constrained by Eq.~\eqref{eq:P2w}, where the nonzero entries of $\chiq$ are given by  Eq.~\eqref{eq:nzchiq}. Disregarding any magnetic effect, the emitted (reflected)  electric field $\bm{E}_r$ due to $\bm{P}$ can be written as~\cite{Sipe1987GF}: 
    \begin{equation}
       \bm{E}_r=\bm{E}_s+\bm{E}_p+f(z)\bm{P},
    \end{equation}
where $\bm{E}_s$ and $\bm{E}_p$ are $s$- and $p$-polarized fields, respectively, and proportional to the incident field. The emitting surface is taken at the $xy$ plane and $f(z)$ is a function giving the contribution of the material layers. Thus, the reflected field is constrained by the symmetry of $\chiq$ in mostly the way $\bm{P}$ is. As the reflected intensity varies as $|\bm{E}_p|^2\propto |\bm{P}|^2$, the crystal symmetry can be identified in the angular distribution of the emitted signal.

\section{Results and discussion}\label{sec7}
Liu {\it et al}~\cite{Liu2013} have conceived a symmetry-based three-band model with nearest and up to third-nearest neighbors, NN and TNN respectively. The NN approximation reproduces {\it ab initio} results in a small vicinity of the gap edges only, and since integrals over the entire BZ are essential for the optical response, in this work we will use the TNN model, which better matches the first principles energy bands across the entire BZ. Details on the TNN Hamiltonian are given in Appendix~\ref{liuTB}, along with the changes introduced by SOC.
 The TB parameters of the whole family of {\it 2H} monolayer TMDCs    were fit at the high symmetry $K$ points to correctly describe the energy gap, so there is no need to introduce corrections by the overlap matrix or a scissors operator~\cite{Cabellos2009}.  From the rest of the paper we will work with WS$_2$, whose TB parameters are given in Table~\ref{tab:table1}, although the same program can be carried out for all other monolayer TMDCs with minimal modifications.  
\begin{figure}[b!]
	\centering
	\includegraphics[width=0.95\columnwidth]{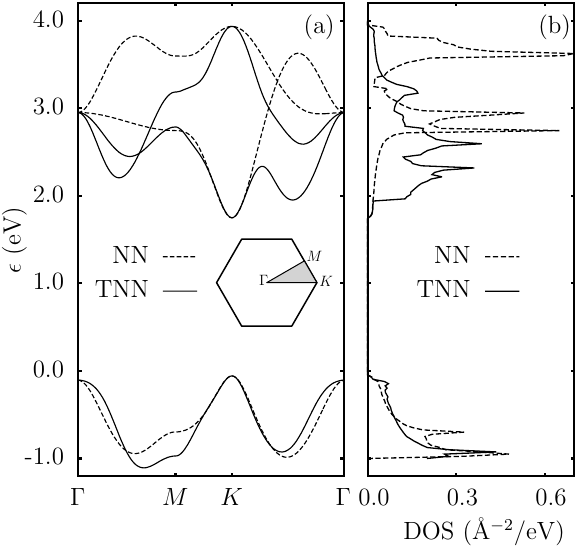}
	\caption{(a) Energy bands of monolayer WS$_2$ in the nearest (NN, dashed line) and third nearest (TNN, solid line) neighbors approximation with TB parameters from~\cite{Liu2013}, along the path $\Gamma\rightarrow M\rightarrow K\rightarrow\Gamma$ in the BZ.  (b) Density of states corresponding to the bands in (a), calculated using the 2D linear tetrahedral method.\label{bandsNN}}
\end{figure}

In using Eqs.~\eqref{chilibz} and~\eqref{chiqibz}, we need to sample the 2D tIBZ in Fig.~\ref{crystal}(d). For simplicity, we use a regular grid of parallelograms parametrized as:
   \begin{equation}
       \bm{k}=(j_x-1)\bm{u}_1/n_1+(j_y-1)\bm{u}_2/n_2,
   \end{equation}
subject to the condition that $\bm{k}$ lies inside the tIBZ and with $j_x$, $j_y$ integers. Vectors $\bm{u}_1$ and $\bm{u}_2$ are depicted in Fig.~\ref{crystal}(c), and $n_2=\lfloor\sqrt{3}n_1/2\rfloor$ to ensure a regular grid ($\lfloor\; \rfloor$ is the floor function). We used $n_1=100$, which results in a sampling of 4393 $k-$points. (As an extra check, we performed calculations with $n_1=150$, equivalent to 9815 $k-$points, which almost exactly match the results with $n_1=100$ when all other parameters are kept the same.)

The imaginary part of the optical susceptibility was integrated by means of a Gaussian broadening approach. The Dirac delta is written as $\delta(\e_{cv}-s\hw)=\delta(w\times(\e_{cv}-s\hw)/w)=(1/w)\delta((\e_{cv}-s\hw)/w)$, where $s=1,2$ [see Eqs.~\eqref{shg11} and~\eqref{shg22}], and $w$ is a  positive energy making the argument in the Dirac delta dimensionless and giving the broadening of it. We use the following approximation~\cite{Methfessel1989}:
\begin{equation}
	\delta(x) \simeq e^{-x^2} \sum_{n=0}^N A_n H_{2n}(x),
\end{equation}
where $A_n=(-1)^n/(n!\, 4^n\sqrt{\pi})$ and $H_{2n}(x)$ are the Hermite polynomials of even order. We have verified that setting $w=0.08\,$eV and $N=3$, convergence in both $\chil$ and $\chiq$ is achieved (except maybe at very sharp peaks, that are more sensitive to $w$). Also, because of the smallness of our system, full BZ integrations were performed, which were in agreement with the relations in Eq.~\eqref{chilibz} and~\eqref{chiqibz}, within a small degree of numerical noise. The real part of both $\chil$ and $\chiq$ is calculated from the imaginary one by means of the Kramers-Kronig relations~\cite{CallawayQTOTSS}:
\begin{equation}
	\text{Re}[\chil(\w)]=\frac{2}{\pi}\mathcal{P}\int_{0}^{+\infty}\frac{\w'd\w'}{\w'^2-\w^2} \text{Im}[\chil(\w')].
\end{equation}
(The symbol $\mathcal{P}$ indicates that the integral is to be evaluated as its Cauchy principal value~\cite{RileyMMPE}). We now present our results separating the cases with zero SOC and full SOC calculations.
\begin{figure}[b!]
	\centering
	\includegraphics[width=0.95\columnwidth]{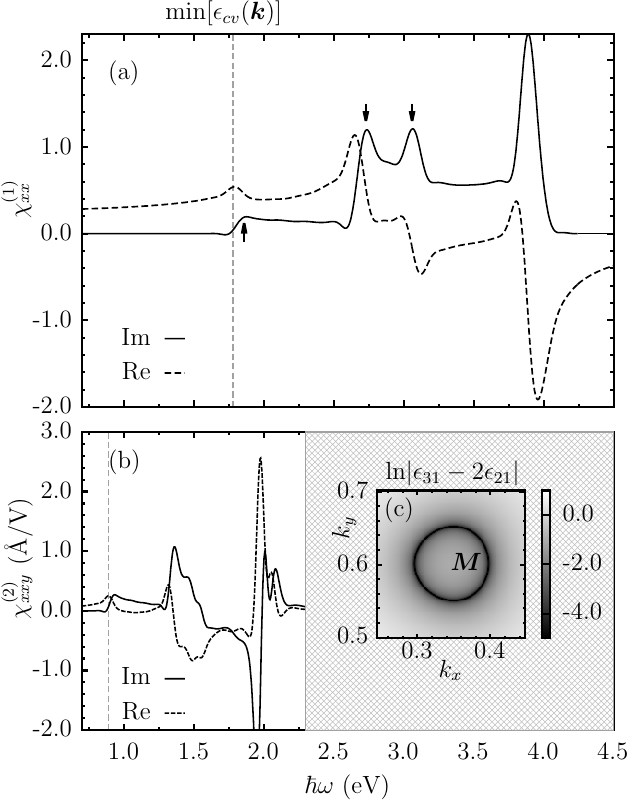}
	\caption{(a) Imaginary and real parts of the $\chi^{\scp (1)}_{xx}$ component of monolayer WS$_2$ as a function of the photon energy $\hw$. (b) Same as in (a) but for the $\chi^{\scp (2)}_{xxy}$ of the nonlinear susceptibility. For $\chi^{\scp (2)}_{xxy}$, photon energies up to 2.3 eV has been included only. (c) Logarithm of the energy difference $\text{ln}|\e_{31}-2\e_{21}|$ showing that it vanishes in a almost circular ring around $M$ point. \label{fig:chi}}
\end{figure}

\subsection{Zero SOC}
 We start describing our results disregarding SOC. The energy bands obtained from the diagonalization of $H_0$ in Eq.~\eqref{TNNh} are shown in Fig.~\ref{bandsNN}(a) along with the ones with nearest neighbors, for the sake of comparison. As it can be seen, both NN and TNN approximations coincide in a very small vicinity of the $K$ point, although they sharply diverge away from it. As shown in~\cite{Liu2013}, the TNN approximation describes well the topmost valence and two bottommost conduction bands throughout the BZ. Since optical excitations occur mostly in this energy region, a model of the optical susceptibility in this energy range is justified. Figure~\ref{bandsNN}(b) depicts the density of states for both the NN and TNN approximations as calculated using a 2D modification of the linear tetrahedral method~\cite{Lehmann1972}.

Figure~\ref{fig:chi}(a) shows the imaginary (solid line) and real (dashed line) parts of the only independent component of the linear optical susceptibility, $\chi^{\scp (1)}_{xx}$. Because WS$_2$ monolayer has a direct band gap, the onset of $\text{Im}[\chil]$ appears for energies greater than the gap size, $\text{min}[\e_{cv}(\bm{k})]=1.81\,$eV,  indicated by the vertical dotted line in Fig.~\ref{fig:chi}(a). The local maxima at energies roughly equal to $\hw=1.86\,$, $2.73\,$ and $3.06\,$eV (indicated by arrows) have been reported before using {\it ab initio} techniques with the projector augmented-wave method~\cite{Wang2015}. However, the sharp maximum around $\hw=4\,$eV has been shown to the smaller than the ones around $2.73\,$eV and $3.06\,$eV. This discrepancy might have its origin in the small number of bands we are using: since $\chil$ depends on $\delta(\e_{cv}-\hw)$, the higher the photon energy $\hw$, the more conductions bands are needed to match the condition $\e_{cv}=\hw$. In this regard, we can say that the three-band approximation gives reliable results up to, say, 3.5 eV. These results differ from those obtained when including many-body effects most importantly in the absence of excitonic peaks below the band gap, as they appear when solving the Bethe-Salpeter equation~\cite{Chaves2017}, and as it is observed in reflectivity measurements~\cite{Li2014}. 

In Fig.~\ref{fig:chi}(b) we plot the imaginary and real parts of $\chi^{\scp (2)}_{xxy}$ [see Eq.~\eqref{chiqibz}]. Our results are presented  in the energy range $0.7\,\text{eV}\moi\hw\moi2.3\,\text{eV}$ only. Because the $\delta(\e_{cv}-2\hw)$ in Eq.~\eqref{shg11}, the onset of the second order susceptibility is now $\text{min}[\e_{cv}(\bm{k})]/2=0.91\,$eV. The contribution $\text{Im}[B_{ijk}]$ from Eq.~\eqref{shg22}, that comprises the resonances with $\hw$, is negligible compared to $\text{Im}[A_{ijk}]$, as it is usually the case. Also, Eq.~\eqref{shg22} presents the phenomenon of {\it double resonance}~\cite{Nastos2005} in the denominator, whereby there is a region in $\bm{k}-$space where $\e_{nv}(\bm{k})=2\e_{cv}(\bm{k})=2\hw$ (for a given $n\neq c$), which makes the entire perturbation expansion inapplicable. In our calculations this happens for $c=2$, $v=1$ and $n=3$. Figure~\ref{fig:chi}(c) is a logarithmic density plot of $|\e_{31}-2\e_{21}|$ in {\it k}-space that shows  that this double resonance appears around the $M$ point. This troublesome term makes the convergence of $\text{Im}[B_{ijk}]$ unattainable, specially for energies just above the threshold $\text{min}[\e_{cv}(\bm{k})]/2$. However, this contribution has been shown to be small and not to affect the nonlinear susceptibility at lower frequencies~\cite{Cabellos2009}. To mitigate this, in the denominators of Eqs.~\eqref{shg11} and~\eqref{shg22} (except, of course, in $\e_{cv}^3$, which is never zero) we added a very small imaginary quantity $\ci\eta$, and then take the real part of it. For Fig.~\ref{fig:chi}(b), $\eta=0.02\,$eV was used. 

Contrary to $\text{Im}[\chil]$, $\text{Im}[\chiq]$ is not thermodynamically required to be positive~\cite{LandauEOCM}. The most prominent features are the peaks at around $\hw=0.94$ and $1.36\,$eV. The sharp dip near $\hw=2\,$eV has been reported before~\cite{Wang2015}, but first principles calculations result in a less deep minimum. Similar to the larger-than-expected peak at $\hw\simeq 4\,$eV for $\text{Im}[\chil]$, the origin of this too large dip is to be found in the limited number of bands used in our calculation. With this, we can state that our results hold good for energies less than $1.75\,$eV, i.e.,  below the spuriously large dip.
\begin{figure}[t!]
    \centering
    \includegraphics[width=0.95\columnwidth]{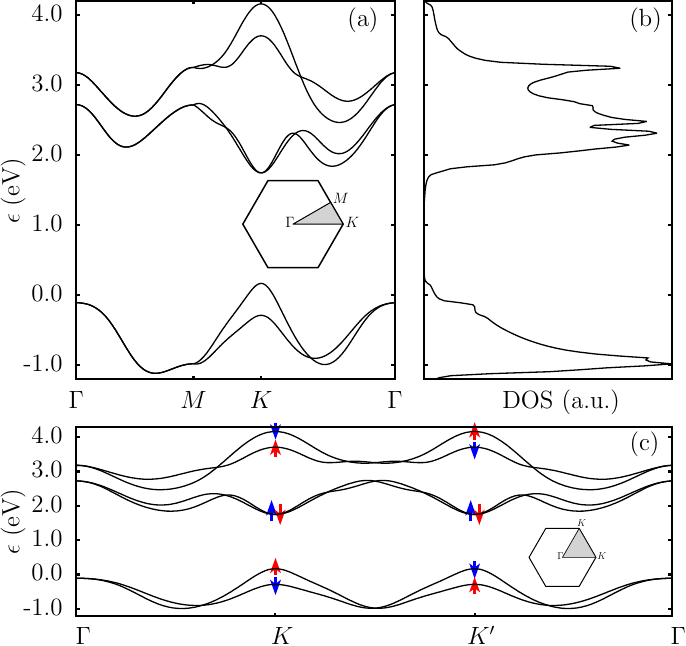}
    \caption{(a) Energy bands with SOC along the path $\Gamma\rightarrow M\rightarrow K\rightarrow \Gamma$ and the associated density of states [(b)]. (c) Energy bands along $\Gamma\rightarrow K\rightarrow K'\rightarrow \Gamma$ showing the spin projections of the Bloch eigenstates at high symmetry points $K$ and $K'$.}
    \label{fig:SOC}
\end{figure}

\subsection{Nonzero SOC}
	Spin-orbit coupling changes these results in a quantitative manner, although the overall profiles and magnitudes of the optical responses are similar. The energy bands along the path $\Gamma\rightarrow K\rightarrow M\rightarrow\Gamma$ and density of states [as obtained using Gaussian broadening with the Dirac delta $\delta(x)$ replaced with $(\eta/\pi)/(x^2+\eta^2)$] are shown in Figs.~\ref{fig:SOC}(a) and (b), respectively. Figure~\ref{fig:SOC}(c) depicts the bands along the path $\Gamma\rightarrow K\rightarrow K'\rightarrow\Gamma$ to highlight the opposite spin content (as indicated by red and blue arrows) at high symmetry points $K$ and $K'$. This is obtained by diagonalizing the Hamiltonian in Eq.~\eqref{eq:HSOC} at the $K$ and $K'$, which can be done in an analytical way. This is also related to the orbital symmetry under threefold rotations of the Bloch eigenstates at these points, and it is responsible for the selective population of regions in {\it k}-space by circularly polarized light~\cite{Xiao2007,Xiao2012}.

\begin{figure}[!t]
    \centering
    \includegraphics[width=0.95\columnwidth]{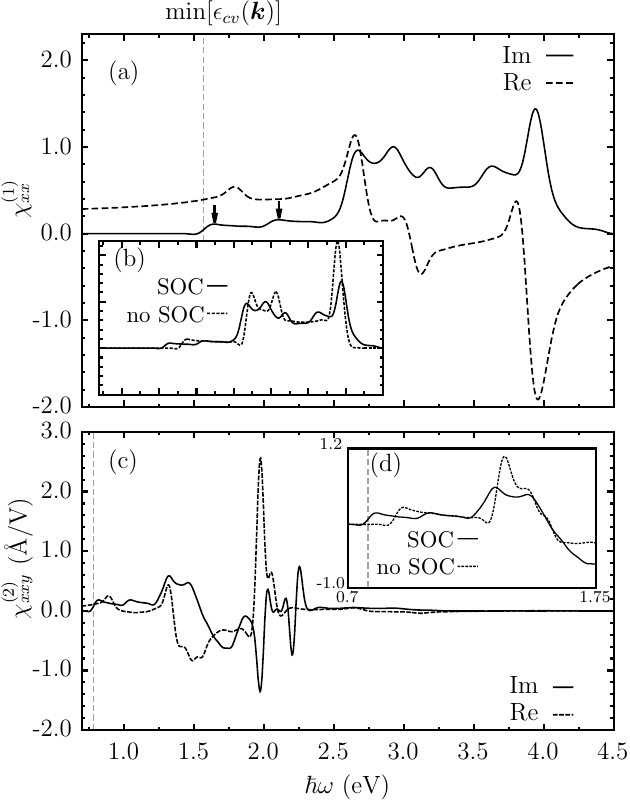}
    \caption{(a) Imaginary (solid line) and real (dashed line) parts of the linear optical susceptibility $\chi^{(1)}_{xx}$ with SOC. (b) Comparison between the imaginary parts of $\chi^{\scp (1)}_{xx}$ with SOC (solid line) and without it (dashed line). Both the horizontal and vertical ranges are the same as in (a). (c) Similar to (a) but for the $xxy$ component of the second order optical susceptibility. Any other parameters are the same as in Fig.~\ref{fig:chi}. (d) Comparison between $\text{Im}[\chi^{(2)}_{xxy}]$ with SOC (solid line) and without it (dashed line). The units are the same as in the main plot.}
    \label{fig:chiSOC}
\end{figure}

    The linear and second order optical susceptibilities are shown in Figs.~\ref{fig:chiSOC}(a) and (c), respectively. The onset of $\text{Im}[\chil]$ is now at $1.57\,$eV, the energy gap with SOC, indicated by a vertical grey dashed line in Fig.~\ref{fig:chiSOC}(a), while the onset of $\text{Im}[\chiq]$ is at $1.57\,\text{eV}/2=0.785\,$eV. For $\chil$, the most visible difference is the appearance of two shoulders [indicated by small black arrows in Fig.~\ref{fig:chiSOC}(a)] where the zero SOC case only shows a single plateau. This is more easily observed in the inset Fig.~\ref{fig:chiSOC}(b), where the imaginary parts of $\chi^{\scp (1)}_{xx}$ with SOC and without it are shown superimposed. As it can be seen from it, SOC does not alter the magnitude of $\text{Im}[\chi^{\scp (1)}_{xx}]$ in an important way. Moreover, the photon energies $\hw$ at which these two features appear correspond to the spin-conserving transitions between the spin-split valence bands and the spin-degenerate lowest conduction band [see Fig.~\ref{fig:SOC}(c)]. Spin conservation comes from the absence of a magnetic field because of our choice of gauge for $\bm{A}(t)$. A similar double-shoulder profile is observed in Figs.~\ref{fig:chiSOC}(c) for $\text{Im}[\chi^{\scp (2)}_{xxy}]$ for photon energies just above the susceptibility onset and below approximately $1.25\,$eV. As it can be seen in the inset Fig.~\ref{fig:chiSOC}(d), which shows $\text{Im}[\chi^{\scp (2)}_{xxy}]$ for the SOC and zero SOC cases, SOC suppresses the sharp maximum just below $1.4\,$eV and splits it into  two local maxima of almost the same height. That SOC creates such a visible splittings in local maxima and minima in the optical susceptibility is a consequence of the large band splittings induced by it. Thus, SOC is an important component in the optical response of monolayer TMDCs and should be included in a correct description of it.
  
\section{Conclusions}\label{sec8}
We showed that in a minimal model of a monolayer TMDC, where the Bloch eigenstates are expressed as combinations of {\it d}-orbitals only, the linear and second order optical susceptibilities can be obtained with reasonable accuracy, as compared with plane-waves {\it ab initio} models. This is more evident at lower frequencies, where most optical experiments are carried out. The key element here is that the TB Hamiltonian written in this approximation matches well DFT results, and that the Bloch eigenstates around the energy gap have an almost complete {\it d}-orbital content, something  that leads to Bloch wavefunction and momentum matrix elements (and, ultimately, optical response) close to the ones obtained using the full DFT wavefunctions. This calculation takes full advantage of the symmetry of the crystal, in a manner more explicitly displayed as compared to other references.

\section{Acknowledgments}\label{sec9}
This project started as a byproduct of another, larger project on nonlinear optics funded by the Department of Energy. I would like to express my gratitude to Luis Enrique Rosas-Hernández for providing me with the SIESTA PAOs of tungsten.

\appendix
\section{Liu's three-band model for monolayer TMDCs }\label{liuTB}
Liu {\it et al}~\cite{Liu2013} have set up a minimal TB model for the whole family of $2H$ monolayer TMDCs  including only the transition metal $d$ orbitals that are even under the $\sigma_h$ mirror operation ($d_{z^2}$, $d_{xy}$ and $d_{x^2}$). When including hopping terms up to third nearest neighbors, the resulting TB Hamiltonian reproduces well the uppermost valence and two lowest conduction bands over the entire BZ. In the basis of Bloch sums given by Eq.~\eqref{sums}, this Hamiltonian can be written as:
\begin{equation}\label{TNNh}
	H_0=\left(\begin{array}{ccc}
		h_{11}   & h_{12}   & h_{13} \\
		h_{12}^* & h_{22}   & h_{23} \\
		h_{13}^* & h_{23}^* & h_{33}
	\end{array}\right).
\end{equation}
In the third-nearest neighbor approximation, the $D_{3h}$ point group of the material constrains the number of independent hopping integrals to just nineteen, indicated in Table~\ref{tab:table1}. By defining $\alpha=ak_x/2$ and $\beta=a\sqrt{3}k_y/2$ [$a$ is the lattice constant, see Fig.~\ref{crystal}(a)], the matrix elements of Eq.~\eqref{TNNh} are given by:

\begin{widetext}
\begin{eqnarray}
h_{11} &= & \varepsilon_1 + 2t_0(2\cos\alpha\cos\beta+\cos2\beta)
 + 2r_0(2\cos3\alpha\cos\beta+\cos2\beta)               
 + 2u_0(2\cos2\alpha\cos2\beta+\cos4\alpha)\,,                \\
h_{22} &= & \varepsilon_2 + (t_{11}+3t_{22})\cos\alpha\cos\beta + 2t_{11}\cos2\alpha 
 +4r_{11}\cos3\alpha\cos\beta + 2(r_{11}+\sqrt{3}r_{12})\cos2\beta \notag\\
 &&+ (u_{11}+3u_{22})\cos2\alpha\cos2\beta + 2u_{11}\cos4\alpha\,, \\
h_{33} &= & \varepsilon_2 + (3t_{11}+t_{22})\cos\alpha\cos\beta + 2t_{22}\cos2\alpha     
       + 2r_{11}(2\cos3\alpha\cos\beta+\cos2\beta)                             \notag \\      
       &&+\frac{2}{\sqrt{3}}r_{12}(4\cos3\alpha\cos\beta-\cos2\beta)             
       + (3u_{11}+u_{22}) \cos2\alpha\cos2\beta + 2u_{22}\cos4\alpha\,,\\            
h_{12} &=& -2\sqrt{3}t_2\sin\alpha\sin\beta + 2(r_1+r_2)\sin3\alpha\sin\beta             
          -2\sqrt{3}u_2\sin2\alpha\sin2\beta                                           \notag\\
          &&+2it_1\sin\alpha(2\cos\alpha+\cos\beta)+2i(r_1-r_2)\sin3\alpha\cos\beta      
          +2iu_1\sin2\alpha(2\cos2\alpha+\cos2\beta)                                  \,, %\\
          \end{eqnarray}
\begin{eqnarray} 
h_{13} &=& 2t_2(\cos2\alpha-\cos\alpha\cos\beta)                                         
        -\frac{2}{\sqrt{3}} (r_1+r_2)(\cos3\alpha\cos\beta-\cos2\beta)
        +2i\sqrt{3}u_1\cos2\alpha\sin2\beta                   \notag\\
       && +2u_2(\cos4\alpha-\cos2\alpha\cos2\beta) +2i\sqrt{3}t_1\cos\alpha\sin\beta     
       +\frac{2i}{\sqrt{3}}(r_1-r_2)\sin\beta(\cos3\alpha+2\cos\beta)                  \,, \\
h_{23} &=& \sqrt{3}(t_{22}-t_{11}) \sin\alpha\sin\beta + 4r_{12} \sin3\alpha\sin\beta    
       +\sqrt{3}(u_{22}-u_{11})\sin2\alpha\sin2\beta                                  \notag\\
       && +4it_{12} \sin\alpha(\cos\alpha-\cos\beta)                                     
        +4iu_{12}\sin2\alpha(\cos2\alpha-\cos2\beta)  .                                 
\end{eqnarray}
\end{widetext}

Spin-orbit coupling is included locally and for the transition metal atom only. Because of the $\sigma_h$ symmetry of the basis $\{d_{z^2},d_{xy},d_{x^2-y^2}\}$, only the term $S_zL_z$ is kept. The final Hamiltonian $H_\text{SOC}$ including SOC in the basis $\{d_{z^2\uparrow},d_{xy\uparrow}$, $d_{x^2-y^2\uparrow}, d_{z^2\downarrow},d_{xy\downarrow}$, $d_{x^2-y^2\downarrow} \}$ is:
\begin{equation}\label{eq:HSOC}
    H_\text{SOC}(\bm{k})=\left[ 
    \begin{array}{cc}
        H(\bm{k}) + \lambda L_z/2 & 0\\
        0 & H(\bm{k}) - \lambda L_z/2
    \end{array}\right],
\end{equation}
where $\lambda$ is a parameter giving the splitting of the bands due to SOC. For WS$_2$, Liu {\it et al} obtained the value $\lambda=0.211\,$eV.

\begin{table}[tb]
\caption{Tight-binding parameters from~\cite{Liu2013} for WS$_2$ with a generalized gradient approximation (GGA). All values are in eV.\label{tab:table1}}
\begin{tabular}{cccccccc}
\hline\hline
$t_0$ & $t_1$ & $t_2$ & $t_{11}$ & $t_{12}$ & $t_{22}$\\ 
$-0.175$ & $-0.090$ & $0.611$ & $0.043$ & $0.181$ & $0.008$ \\
\hline
 $r_0$ & $r_1$ & $r_2$ & $r_{11}$ & $r_{12}$ & $u_0$ \\ 
 $0.075$ & $-0.282$ & $0.356$ & $2.015$ & $2.014$  & $2.056$ \\
\hline
$u_1$ & $u_2$ & $u_{11}$ & $u_{12}$ & $u_{22}$ & \\
$2.045$ & $0.659$ & $3.014$ & $0.457$ & $0.478$ & \\
\hline
$\varepsilon1$ & $\varepsilon_2$  \\
$0.717$ & $1.916$  \\
\hline
\hline
\end{tabular}
\end{table}


%\bibliography{references}

\begin{thebibliography}{58}
\bibitem [{mis()}]{misc1}%
  \BibitemOpen
  \href@noop {} {}\bibinfo {note} {The optical susceptibility is obtained from the macroscopic induced polarization, which in turn comes from the induced current. The latter is proportional to the electron velocity through the corresponding operator. Without SOC, this velocity operator is trivially related to the momentum operator, namely $\bm{v}=\bm{p}/m$. Spin-orbit coupling changes this, and thus $\bm{v}$ has to determined from its very definition $\bm{v}=(1/i\hbar)[\bm{r},H]$.}\BibitemShut {Stop}%
\end{thebibliography}

\end{document}